\documentstyle[editedvolume,numreferences]{crckapb} 
\input{epsf}

\newcommand{\bq}{\begin{equation}}
\newcommand{\ee}{\end{equation}}
\newcommand{\fr}[2]{\frac{#1}{#2}}
\newcommand{\eps}{\varepsilon}

\begin{opening}
\title{COULOMB BLOCKADE IN QUANTUM DOTS\protect\\ WITH
OVERLAPPING RESONANCES:} 
\subtitle{Towards an Explanation of the Phase Behaviour in the
Mesoscopic Double-Slit Experiment}

\author{P.G. SILVESTROV$^{1,2}$}
\institute{$^{1}$Budker Institute of Nuclear Physics,\\ 630090
Novosibirsk, Russia}
\author{Y. IMRY$^2$}
\institute{$^{2}$Weizmann Institute of Science,\\ Rehovot 76100, Israel}

\end{opening}

\runningtitle{COULOMB BLOCKADE WITH OVERLAPPING RESONANCES}

\begin{document}

\begin{abstract}

Coulomb blockade (CB) in a quantum dot (QD) with one anomalously
broad level is considered.
In this case many consecutive pronounced CB peaks correspond to
occupation of one and the same broad level. Between the peaks
the electron jumps from this level to one of the narrow levels
and the transmission through the dot at the next resonance
essentially repeats that at the previous one. This offers a
natural explanation to the recently observed behavior of the
transmission phase in an interferometer with a QD. 
Single particle resonances of very different width are natural
if the dot is not fully chaotic. This idea is illustrated by the
numerical simulations for a non-integrable QD whose classical
dynamics is intermediate between integrable and chaotic.
Possible manifestations for the Kondo experiments in the QD are
discussed.

\end{abstract}

\section{Introduction. The Double-Slit Experiment}\label{sec:1} 

Much progress has recently been achieved in the fabrication and
experimental investigation of ultrasmall few-electron devices - such
as Quantum dots \cite{LKCM}. A useful theoretical
tool for the investigation of chaotic QD-s is the Random Matrix theory
(see reviews \cite{Beenakker,Weidenmuller}). Nevertheless, many
experimentally observed features of these artificial multi-electron
systems still have not found a reasonable theoretical
explanation. 

A challenging problem which has resisted adequate theoretical
interpretation arises from the experiments \cite{Yacoby,Heiblum}
which determine the phase of the wave transmitted through the
QD. The goal of this talk will be to present a mechanism which
may lead to a satisfactory explanation of these results.
The QD in the experiment was imbedded into
one arm of the Aharonov--Bohm (AB) interferometer, which allowed
to measure not only the magnitude of the transmission, but also
the phase acquired by the electrons traversing the dot. The
total current through the interferometer is~\cite{Heiblum} 
\bq\label{Jtot}
J_{total}\sim |t_{QD} +e^{i\phi_{AB}}t_{sl}|^2 \ ,
\ee
where $t_{QD}$ and $t_{sl}$ are  the transmission amplitudes through the
QD and the
second reference arm of the interferometer. The AB phase is
proportional to the magnetic flux $\Phi$ threading the
interferometer $\phi_{AB}= 2\pi \Phi/\Phi_0$. The {\it complex}
amplitude $t_{QD}$ (as well as the number of electrons in the
dot) is slowly changed by the plunger gate voltage $V_g$ and
does not depend on the weak magnetic field. Parametrising
the amplitudes as $t=|t|\exp(i\theta)$ the interference term in
eq.~(\ref{Jtot}) becomes, in obvious notation:
\bq\label{int}
|t_{QD}| |t_{sl}| \cos (\phi_{AB}+\theta_{sl}-\theta_{QD}) \ .
\ee
Thus by independently changing  $V_g$ and $\phi_{AB}$ one may
find both $|t_{QD}|$ and the variation of the phase $\Delta
\theta_{QD}$ as a function of the gate voltage.

In the experiments \cite{Yacoby,Heiblum} the
coherent component of the electron transport through the QD was
measured directly for the first time. 
Later experiments allowed also an investigation of the dephasing
rate of the electron state in the QD~\cite{Buks}. Here we
discuss the unexpected and not understood yet observed behavior of the phase
$\theta_{QD}$ of the transmitted electron.

\begin{figure}
\epsfxsize=10cm
\epsffile{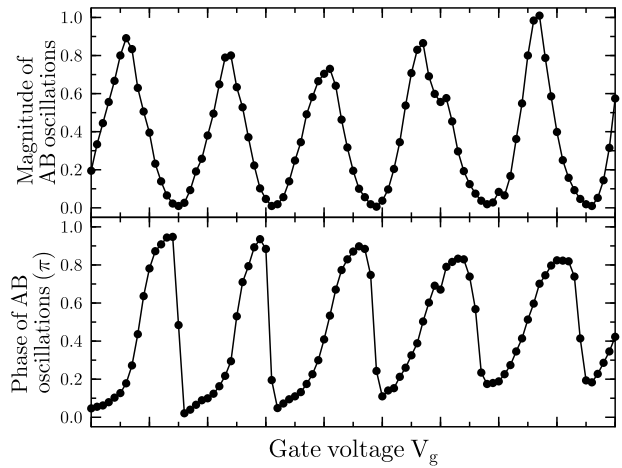}
\caption{The magnitude of AB oscillations $\sim |t_{QD}|$
(arbtrary units)
and the variation of the phase $\Delta \theta_{QD}$ (in units of
$\pi$) as a function of the gate voltage.}
\end{figure}

This unusual behavior of the phase is illustrated in fig.~1
(an approximate drawing, very close to the real
experimental figure of ref.~\cite{Heiblum}). As one can see from
the figure, in accordance with the Breit-Wigner picture, the
measured phase increased by $\pi$ around each CB peak.
Absolutely unexpected, was a fast jump of the phase by $-\pi$
between the resonances just at the minimum of the transmitted
current. Such a behavior is in evident contradiction with the theoretical
expection for transmission  via
the consecutive levels in 1-dimensional quantum well. The phase
shift between the resonances in this case simply accounts for
the number of nodes of the (almost real) wave function inside
the dot, which is changed by one in $1d$ at the consecutive
resonances.

A number of theoretical papers attempted to find an explanation
of the $-\pi$ jumps. It is relatively easy to find a mechanism
which leads to the fast drops of phase in some part of the CB
valleys. It is more complicated to explain why for the
long series of peaks the increase of the phase by $\pi$ at the
resonance will be accompanied by the $-\pi$ drop at the valley.
In two--dimensional QD the phase drops associated with the nodes
of the transmission amplitude arise already within the
single--particle picture (the Fano mechanism). However, in order
to have a sequence of such events one should consider a QD of a
very special form~\cite{Deo,Fano}. For electrons with spin and
only one level in the QD, the phase drops between the resonances
corresponding to adding of the first and second electron to the
dot~\cite{Oreg}, but again this effect is not easy to generalize
for a series of peaks. The mechanism of
refs.~\cite{Hackenbroich,Gefen} makes nontrivial assumptions
on the geometry of the QD and the way it changes under the
change of plunger gate voltage. An interesting generic mechanism, which
indeed may lead to the sequence of $\sim U_{CB}/\Delta$ drops
between the resonances was suggested in ref.~\cite{Baltin}.
However, within this approach the extra phase jumps do not occur
in the same fashion in consecutive resonances.

In this paper we propose a mechanism according to which the
transmission at many CB peaks proceeds through one and the same
level in the QD. This means that the phases at the wings of
different resonances should coincide and therefore the increase
by $\pi$ at the resonance must be compensated. This compensation
occurs via narrow jumps between the resonances and is
accompanied by a fast rearrangement of the electrons in the dot.

The experiment~\cite{Heiblum} was clearly done in the CB regime.
However, in order to be able to measure the AB oscillations in
the CB valleys the dot was significantly "opened". As
a result, the widths of the resonances turn out to be rather
large, being only few times smaller than the charging energy
(see fig.~1). Also the widths and heights of all observed
resonances are quite similar. These features of the
experimental results, which have not attracted so wide an attention
as the phase jumps, also  naturally follow from our
mechanism.

It is generally believed that the CB is observed only if the
widths of resonances are small compared to the single-particle
level spacing in the dot $\Delta$.  This condition assumes that
couplings of all levels to the leads are of the same order of
magnitude. However, as we will show, even for nonintegrable
ballistic QD-s the widths of the resonances may vary by orders
of magnitude. In this case it does not make sense to compare the
width of few broad resonances with the level spacing, determined
by the majority of narrow, practically decoupled, levels. Our
goal in this paper is twofold. First, in the following section
we will consider a simple numerical example which shows that
coexistence of narrow and broad resonances is quite natural in
ballistic QD-s with relatively strong coupling to the leads.
Second, we will consider how the CB and filling of the levels in
the dot are changed in the presence of a dominant single broad
resonance.

A useful theoretical model for the description of charging
effects in QD-s is the tunneling Hamiltonian (see
e.g.~\cite{CB}) 
\begin{eqnarray}\label{Ham}
&\,&H=\sum_i \eps_i a^+_i a_i + U_{CB}(\sum_i a^+_i a_i-N_0)^2/2  
\\ &\,& \ \ \ \   +   \sum_k\eps^L_k b^{L+}_k b^L_k +
\sum_{k,j}[t^L_j a^+_j b^L_k +h.c.]
+ L\leftrightarrow R
\ . \nonumber
\end{eqnarray}
Here $a(a^+)$ and $b(b^+)$ are the annihilation(creation)
operators for electron in the dot and in the lead, $L$ and $R$
stand for the "left" and "right" lead and summation over spin
orientations is included everywhere. 

Refined theoretical treatment of the charging effect within the 
Hamiltonian~(\ref{Ham}) includes modern theoretical tools such
as bosonisation and mapping onto the Kondo problem
\cite{Matveev}. For our purposes, it will be possible to
simplify further eq.~(\ref{Ham}). We will consider
the case of only one level $N$ in the dot significantly coupled 
to the leads (only $t^{L,R}_N\ne 0$). If in addition the width
of this level is larger than the single-particle level spacing
$\Delta$, a very nontrivial regime of CB may be described by
means of simple second order of perturbation theory estimates.
Surprisingly this simple limit of CB in the QD with broad level
have not been considered yet.

\section{Semi-Chaotic Quantum Dots}\label{sec:2}

\begin{figure}
\epsfxsize=11.5cm
\epsffile{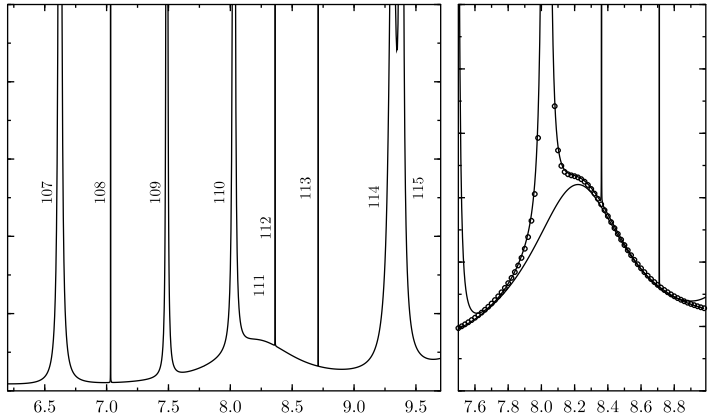}
\caption{The energy dependence of the probability to find the
electron in the QD for fixed incoming(outgoing) current. The
numbers of the corresponding levels in the dot are shown. Right
panel: A part of figure close to the resonance $\# 111$ together
with the fit by two Breit-Wigner peaks (circles) and single
broad peak with $\Gamma = 0.7$.}
\end{figure}

In this section we will show by explicit numerical
example how the anomalously broad levels may appear even
in sufficiently large and irregular QD.

\begin{figure}
\epsfxsize=10cm
\epsffile{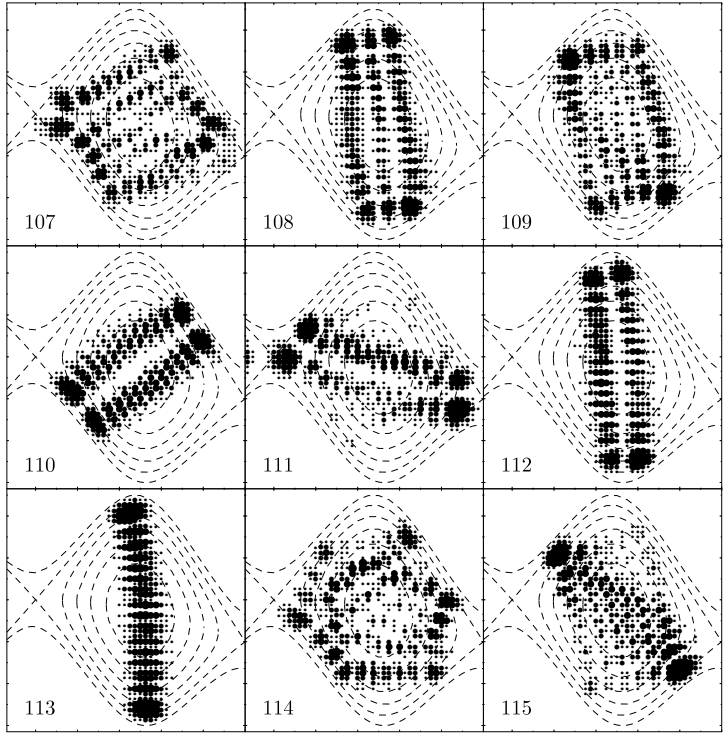}
\caption{The density ($\sim$dark area) of electrons in the dot for 
the nine levels
shown in fig.~1. 
Dashed lines are the lines of
constant potential 
$V=20,0,-20,-40,-60$ and $-80$.  
The single-channel wire is attached from the left.
About $95\%$ of the norm of the wave function in the dot is shown.}
\end{figure}

A simple example of a system for which the widths differ
drastically is the integrable QD\cite{Hackenbroich,Gefen}.
Integrable QD-s having only few ($2\div 4$) electrons have been
produced recently \cite{Leo}.  However, for large numbers of
electrons $N_e\sim 100 \div 1000$ it will be much more difficult
to produce an integrable QD\footnote{It is not clear how
close to integrable was the QD used in the
experiment\cite{Heiblum}. However, the QD containing $\sim 200$
electrons 
was $\sim 50$ times smaller than the nominal elastic mean free path.
Thus, disorder should not be essential for the
dynamics of the electrons.}. Nevertheless, at least in classical
mechanics, one finds a considerable gap left between integrable
and fully chaotic systems. Even in the nonintegrable dot there
may coexist two kinds of trajectories - quasi-periodic and
chaotic. In this case, in 2-dimensions any trajectory (even the
chaotic one!) does not cover all the phase space formally
allowed by energy conservation.  Consequently, the corresponding
wave functions do not cover all the area of the QD. If such a
regime is or may be realized in QD-s, it easily explains, why
the widths of the resonances may vary by orders of magnitude.
Moreover, many other features of such a QD may differ strongly
from those of the chaotic QD (CB intervals, energy level
statistics\cite{Stopa}). To {\it illustrate} this idea we have
performed numerical simulations for a model QD of the size $l$
coupled smoothly to the leads with the potential
\bq\label{v} 
V= -
x^2\bigl(1-{x}/{l}\bigr)^2 + \bigl(y+0.2 {x^2}/{l}\bigr)^2
\Bigl(1+2 \bigl(2{x}/{l}-1\bigr)^2 \Bigr) .  
\ee 
The dot is strongly nonintegrable, but, similarly to the
experimental geometry~\cite{Heiblum}, sufficiently symmetric.
For numerical simulations we considered the QD on the lattice
having a lattice spacing equal to unity and $l=40$. The kinetic
term is given by the standard nearest-neighbor hopping
Hamiltonian. With the hopping matrix element $\tau\approx 40$
our dot has about $100$ electrons ($\sim200$ with spin).  The
calculations were performed for the right lead closed by a hard
wall at $x=l$ (see ref.~\cite{preprint} for some more details).
Only one mode may propagate along the left lead within the
energy interval $5.8<\eps<10.3$ (the energy band is
$V<\eps<V+8\tau$). The phase of the electron in the lead was not
fixed and we were able to find a solution of the Schr\"{o}dinger
equation at any energy. The probability to find an electron in
the dot as a function of energy for fixed in- and out-going
current in the lead is shown in fig.~2. Above-barrier resonances
with very different widths are seen. Moreover, the width of the
resonance $\# 111$ turns out to be few times larger than the
level spacing. The origin of this hierarchy of widths becomes
clear from fig.~3, where we have plotted $|\psi|^2$ in the QD at
these resonances. The quantized version of different variants of
classical motion may be found on this figure. The most narrow
level $\# 113$ corresponds to short transverse periodic orbit.
Other broader levels, such as $\# 109$, may be considered as the
projections of the invariant tori corresponding to
quasi--periodic classical motion. This classical trajectory
reaches the line $V(x,y)=\eps$ only at few points. The presence
of different types of (quasi)periodic motion is natural for the
nonintegrable dot (see the wave functions $\# 113, \# 110, \#
111$ and less pronounced $\# 115$). The candidates for chaotic
classical motion (e.g. $\# 114$) also correspond to relatively
broad resonances. Even in this case only a part of QD is covered
by the trajectory.

Most interesting is the level $\# 111$, which is well coupled to
the leads. The corresponding classical trajectory covers some
invariant tori in the dot starting from the left contact. The
Breit-Wigner width of this resonance is $\Gamma\approx 0.7$
(same units as in fig.~2), which is sufficiently larger than
$\Delta$.

Essentially the same pictures with "regular" and "chaotic" wave
functions may be found in ref.~\cite{Stopa}, where also the
potential of the QD was treated selfconsistently. Other
mechanisms which may lead to the "stability" of the broad level
will be considered in the next section.

\section{Coulomb Blockade for a Single Broad Level}\label{sec:3}

With our numerical {\it example} we have investigated only the 
single particle properties of the ballistic nonintegrable QD. 
Now let us turn to the many--particle effects. First of all, the
parameters of the dot are slowly changed by the change of
the plunger gate voltage $V_g$. We describe this effect by 
letting the levels in the dot flow as 
\bq\label{1}
\eps_i=\eps_i{(V_g=0)} -V_g \ \ , \ \ i=1,2,3 ... \ .
\ee
The occupied levels are now those with $i\le 0$.
The energies of the electrons in the wire are given by
\bq\label{11}
\eps(k)={k^2}/{2m}-E_F .%=\fr{k^2-k_F^2}{2m} .
\ee
Here $k={n\pi}/{L}$ with large integer $n$ and $L$ is the length
of the wire. As it was mentioned in the introduction, we suppose that
only one ($N$-th) level in the dot is coupled strongly to the
leads. It is enough to consider coupling with only one
lead~\cite{Glazman}. Let the corresponding matrix element be
$t_N$ and 
\bq\label{2}
\Gamma_N = 2\pi |t_N|^2 {dn}/{d\eps} \gg \Delta \ . 
\ee
In our example (after the doubling of number of levels due to
spin) the values were $\Gamma_N/\Delta\sim 5$. The
charging energy is assumed  even larger, $U_{CB}\gg\Gamma$. The
widths of other levels are much smaller than $\Delta$
and may be neglected.  Our aim is to show, that transmission of
a current at about $\Gamma/\Delta$ consecutive CB peaks will
proceed through one and the same level $\eps_N$ in the dot.

Let us start with  spinless electrons. 
As long as we assume $t_i=0$ for all $i\ne N$, the spectrum of the
tunneling Hamiltonian (\ref{Ham}) consists of completely
decoupled branches corresponding to different occupation numbers
of narrow levels. Let at first only the levels with $i\le 0$ be
occupied. It is easy to find the second order correction to the
total energy of the wire and QD, when the $N$-th level lies far
above the Fermi energy ($\eps_N(V_g)\gg \Gamma$)
\bq\label{3}
\Delta E^{(0)}_{tot}=\int_0^{k_F}
\fr{|t_N|^2}{\eps(k)-\eps_N} \fr{L}{\pi} dk
= \fr{- \Gamma}{2\pi} \ln\left( \fr{4E_F}{\eps_N(V_g)} \right) .
\ee
The levels in the wire are lowered due to the repulsion from the
unoccupied level $\eps_N$. We will not include into $E_{tot}$
the trivial constant which arose due to occupation of the
decoupled levels with $i\le 0$ and unperturbed electron energies
in the lead. Generalization of  eq.~(\ref{3}) for the case
of negative $\eps_N$, $\eps_N \ll -\Gamma$ (the broad level being
below the Fermi energy) is straightforward (note that the level
$\eps_N$ is occupied, {\em not the level $\eps_1$ as 
one might expect}):
\bq\label{5}
E^{(0)}_{tot}= \eps_N -\fr{\Gamma}{2\pi} \ln\left(
\fr{4E_F}{|\eps_N|} \right) \, .
\ee
Here the first term, $\eps_N<0$, accounts for the energy loss due to
replacement of one electron from lead to the dot. The correction due
to the second order of perturbation theory now includes both lowering
of levels with $\eps(k)<\eps_N$ and rising of those with
$\eps(k)>\eps_N$. The perturbative treatment fails for
$|\eps(k)-\eps_N|\le \Gamma$, but the corresponding shifts of levels
below and above $\eps_N$ evidently compensate each other, which is
equivalent to taking the principal value of the integral in
Eq~(\ref{3}). 

Finally, the exact solution for a single state interacting with a
continuum is also known (see e.g.~\cite{Bohr}). A precise
treatment of this Breit-Wigner-type situation, along the lines
of ref.~\cite{Landau}, yields for $E_F\gg \eps_N,\Gamma$:
\bq\label{4}
E^{(0)}_{tot} =
\fr{-\Gamma}{4\pi} \left[
\ln\left(\fr{16E_F^2}{\eps_N^2+\Gamma^2/4}\right)
 +2 \right] 
 + \fr{\eps_N}{\pi} 
\cot^{-1}  \fr{2\eps_N}{\Gamma}  ,
\ee
which coincides with Eqs.~(\ref{3},\ref{5}) at
$|\eps_N|\gg\Gamma$. 

Let us now consider the branch where the level $\eps_1$ is 
occupied. The energy of this electron is $\eps_1 (V_g)$. However,
adding one more electron via the hopping $t_N$ {\em now costs
$\eps_N(V_g)+U_{CB}$}. The ensuing reduction of the downward
shift of the level $E^{(1)}_{tot}$ is of crucial importance. The
analog of Eq.~(\ref{3}) for $\eps_N+U_{CB} \gg \Gamma$ now reads
\bq\label{6}
E^{(1)}_{tot}= \eps_1 -\fr{\Gamma}{2\pi} \ln\left(
\fr{4E_F}{\eps_N+U_{CB}} \right) \, .
\ee
The initial energy of the electron gas in the leads as well as
the contribution from $\eps_i$ with $i\le 0$ may be included as
an additive constant in eqs. (\ref{5}~-\ref{6}).

Within $-U_{CB}<\eps_{1,N}<0$, both eqs. (\ref{5}) and (\ref{6})
are valid. Since $\eps_N> \eps_1$ one expects $E^{(1)}_{tot}$
to be the true ground state in this region. However, because 
in our case $\Gamma\gg \Delta$ the second term in
eqs.~(\ref{5},\ref{6}) may invalidate this expectation.
For small $V_g$ one has $E^{(0)}_{tot}<E^{(1)}_{tot}$ and the Eqs.
(\ref{3}~-\ref{4}) describe the true ground state of the 
system. However, the two functions $E^{(0)}_{tot}(V_g)$ and
$E^{(1)}_{tot}(V_g)$ cross at
\bq\label{8}
\eps_N(V_g)=-{U_{CB}}/{[\exp\{ 2\pi(\eps_N -\eps_1)/\Gamma\}
+1]}  
\ee
and the ground state jumps onto the branch $E^{(1)}_{tot}$. The
current-transmitting {\it virtual} state $N$ now again has a
positive energy. Thus, the phase of this transmission has
returned to what it was before the process of filling of state
$N$ and the subsequent sharp jump into the state where level $1$
is filled. It is the latter jump which provides the sharp drop
by $\pi$ of the phase of the transmission amplitude, following
the increase by $\pi$ through the broad resonance. Thus, many
($\sim (\Gamma/\Delta) ln(U_{CB}/\Gamma)$) consecutive resonances
are due to the transition via one and the same level $N$.

For electrons with spin the Breit-Wigner-related formula~(\ref{4})
does not work. One should consider the Anderson impurity model.
However, far from the resonance the perturbation theory may still be
used (we assume that the temperature is large enough to be away from the
Kondo effect [18-20]).
%\cite{Glazman}). 
Eq. (\ref{3}) acquires
only the overall 
factor 2 due to spin $E^{(0)}_{tot} \rightarrow 2E^{(0)}_{tot}$. The
logarithm in Eq.~(\ref{6}) is also multiplied by $2$. Instead of 
Eq. (\ref{5}) one has
\bq\label{14}
E^{(0)}_{tot}= \eps_N -\fr{\Gamma}{2\pi} \left\{ \ln\left(
\fr{4E_F}{|\eps_N|} \right) +\ln\left(
\fr{4E_F}{\eps_N+U_{CB}} \right) \right\} .
\ee
The first logarithm here is the second-order correction due to the
jumps from the occupied orbital in the dot back to the wire.  Another
logarithm accounts for the jumps from the wire onto the second,
unoccupied orbital. The ground state of the system is now doubly
degenerate due to two orientations of spin in the dot.  The result
(\ref{8}) is basically unchanged.

The crossing of  energy levels
$E^{(0)}_{tot}$ (\ref{5}) and $E^{(1)}_{tot}$ (\ref{6}) becomes
avoided
if one introduces the small individual width $\Gamma_1$ of the
narrow level $|1\rangle$, but the phase drop remains sharp
$\delta V_{g} \sim(\eps_1-\eps_N) \sqrt{\Gamma_1/\Gamma_N}$
\cite{preprint}. 

In the simplified model Eq.~(\ref{1}) all levels in the dot are
changed in the same way by the plunger voltage.  Taking into
account the different sensitivities of longitudinal and
transverse modes to the plunger~\cite{Hackenbroich,Gefen} may
allow to keep our broad level $\eps_N$ even longer within the
relevant strip of energy.  This may provide an explanation of
even longer sequences of resonances accompanied by the $-\pi$
jumps.

Generalization of our approach for $N<0$ (still
$|\eps_N|<\Gamma$, the broad level immersed into the sea of
occupied narrow levels) is straightforward.

Also in a more refined approach, adding new electrons into the
QD should cause a slow change of the selfconsistent potential
$V(x,y)$. The total energy of the dot and the wire will be
lowered in the presence of strongly coupled levels.  This may
cause the potential of the QD to automatically adjust to allow
such levels, which will support our explanation of the
experiment of Ref.~\cite{Heiblum}.

Even though we have shown in the section~3 that the broad levels
are natural for the strongly coupled ballistic QD-s, one may
wonder, why it happens that in the experiment such a level was
found just close enough to the Fermi energy. We see two possible
explanations. First, the self-consistent shape of the QD may
indeed be 
automatically adjusted in order to have such a level.
Second,
the parameters of the QD might have been tuned in the course of
preparation of the experiment while opening up the dot.

\section{ Kondo Effect}\label{sec:4} 

Recent experiments
%\cite{Kondo1,Kondo2,Kondo3,Kondo4} 
[23-26]
observed the  increase of the conductance $G$ of an appropriate
QD at low 
temperature in the CB valleys corresponding to the odd number of
electrons, which has stimulated the renewed interest in the Kondo
effect in QD-s [18-20]
%\cite{Glazman}.

The correction to $G$ is due to the spin-flip interaction
of the unpaired electron in the dot with the leads. For example, at
the upper wing of the first charging resonance 
\bq\label{Kondo}
G_{K}=G_0\left( 1+ \fr{const\times \Gamma}{E_F-\eps}\ln\left(
\fr{U_{CB}}{T}\right) \right) \ , \ \Gamma\ll E_F-\eps\ll U_{CB}
\ . 
\ee
Here $\eps$ is the energy of single particle state in the dot
and the unperturbed conductance at the wings of resonance is
$G_0\sim \Gamma_L\Gamma_R/(E_F-\eps)^2$. However, already in the
first experiments,  effects which do not find an explanation
within the straightforward Kondo mechanism were
observed~\cite{Kondo3,Kondo4}. 

\begin{figure}
\epsfxsize=12.5cm%13.2cm%15cm%8.8cm
\epsffile{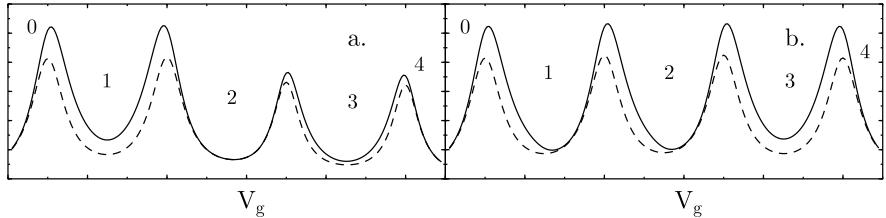}
\caption{The gate voltage dependence of the conductance
(schematic): $a.$~for two narrow resonances ($\Gamma \ll
\Delta$), $b.$~for one broad and one narrow resonance. Dashed
line $T\gg T_K$ (but $\Gamma \gg T$), solid line $T\sim T_K$.
The valley number 
coincides with the number of electrons added to the dot.}
\end{figure}

The correction associated with the spin flip blows up at the
Kondo temperature
$T_K= \sqrt{U_{CB}\Gamma}\exp\left\{
\pi{(\eps-E_F)(\eps+U_{CB}- E_F)}/{2\Gamma U_{CB}}\right\}
$~\cite{Haldane}. 
The natural way to observe the Kondo effect is to
fabricate the devices with large $T_K$. $T_K$ is
large in  very small (up to $\sim 50$ electrons in
real experiments) and sufficiently open QD-s. 

In this section we will consider how the Kondo effect behaves in
the case of overlapping resonances. In fact we do not need now
to have $\Gamma \gg \Delta$. The interesting new effects appear
already if one has two levels $\eps_1$ and $\eps_2$ one narrow
and another broad with $\Gamma \sim \eps_2 - \eps_1$. The
prediction for the Kondo--type correction to the conductance in
this case follows immediately from the consideration of the
previous section without any additional calculation. The result
is shown in fig.~4.

In fig. 4a. we have shown schematically the conductance as
a function of gate voltage following the occupation of two
narrow levels (the usual Kondo effect). The first two similar
peaks correspond to occupation of level $\eps_1$, the other two
peaks correspond to occupation of the second level $\eps_2$,
having slightly different width. At low temperature $T\sim T_K$
the conductance grows up at the odd valleys $1$ and $3$. 

Quite different is the Kondo effect in the case of the broad
peak shown in  fig.~4b. At the left peak the broad level  
crosses the Fermi energy for
the first time. Since $\Gamma\sim
\eps_2 -\eps_1$ no matter which energy is larger,
the broad level is occupied at this resonance. At 
low $T\sim T_K$
the conductance increases at the right wing of the resonance
($1$ electron in the QD) and does not change at the left wing
(no electrons in the QD). Something of interest happens at
the middle of the valley $1$. Here, in accordance with the theory
developed in the previous section, the electron in the dot jumps
from the broad to the narrow level and becomes "invisible" for
the exchange with the leads. Thus though formally the spin of
the QD equals  $s=1/2$, there is no Kondo effect at the left
wing of the second resonance\footnote{Some increase of the
conductance due to spin exchange still may take place here, but
the correction to $G$ is proportional to the small width of
narrow level}. The second peak is associated again with the
population of the broad level by one electron. As for the
valley $1$, one finds the Kondo effect only in the left half of
the second valley. In the middle of this valley the electron
configuration is rearranged again and the second
electron jumps to the narrow level. After that the 
occupied narrow level does not play any role in the
transmission. The usual Kondo effect takes place in the
valley 3 in fig.~4b. 

Thus, our main result of this section is the prediction of
possibility to have Kondo effect in even valleys in the QD.
Remarkably, something very close to this prediction was observed
in the recent experiment in Stutgart~\cite{Kondo3}. The authors
of ref.~\cite{Kondo3} have observed in a few samples the
increase of the zero-bias differential conductance for odd number of
electrons in the dot (such as the valley $3$ on our fig.~4b). In
addition, in two samples they have found the Kondo--type peak in
the preceding valley, which starts at zero bias and is then (with
increase of $V_g$) shifted smoothly to the (small) non-zero
bias voltages. In this paper we consider only the zero bias
conductance, but already the zero bias part of this observation
of ref.~\cite{Kondo3} agrees well with what we have shown on the
valley $2$ of fig.~4b.\footnote{The "elementary event" shown on
the fig.~4b. consists of the sequence of $4$ peaks(valleys), but
ref.~\cite{Kondo3} reports
measurements  for two valleys only}. The explanation of the finite
bias part of the experiment should include
nonequilibrium effects. The work in this direction is in
progress.

\section{Conclusions}\label{sec:5}

To conclude, motivated by the measurements of the transmission
phase in the QD~\cite{Yacoby,Heiblum} we investigated the
charging effects in the QD with a single anomalously broad level.
Contrary to the common expectation the pronounced CB takes place
here even for  $\Gamma\gg\Delta$. In this case upon increasing
$V_g$, it is energetically favorable to first populate in the
dot the level strongly coupled to the leads.  At a somewhat
larger $V_g$ a sharp jump occurs to a state where the "next in
line" narrow level becomes populated. This jump accounts for the
sharp decrease by $\sim \pi$ of the transmission phase observed
in the experiment. The absence of significant fluctuations of
the strength of resonances seen in the experiment~\cite{Heiblum}
and their large widths are clear within our mechanism.  The
current transmission through such QD resembles the behavior of
rare earth elements, whose chemical properties are determined
not by the electrons with highest energy, but by the "strongly
coupled" valence electrons.
An explicit numerical example shows how such broad levels may
appear in ballistic (non--integrable) QD-s.

The overlapping of single-particle resonances may take place
also in the Kondo experiments in QD-s, where in order to
increase the Kondo temperature the dot is usually sufficiently
opened. New low-temperature effects take place in this case,
including the Kondo effect in even valleys. This may provide an
explanation of the unexpected results of recent
experiments~\cite{Kondo3,Kondo4}.

\section{Acknowledgements}

The work of PGS was
supported by RFBR, grant 98-02-17905. Work at WIS was supported by
the Albert Einstein Minerva Center for Theoretical Physics and by
grants from the German-Israeli Foundation (GIF), the Bond fund 
and the Israel Science Foundation, Jerusalem.


\begin{thebibliography}{99}

\bibitem{LKCM} L.~P.~Kouwenhoven {\it et al., Mesosopic Electron
Transport, Proceedings of the NATO ASI,}
edited by L.~L.~Sohn, L.~P.~Kouwenhoven and G.~Sch{\"o}n
(Kluwer 1997).

\bibitem{Beenakker}  C.~W.~J.~Beenakker, Rev. Mod. Phys. {\bf
69}, 731 (1997).

\bibitem{Weidenmuller} T.~Guhr, A.~M\"uller-Groeling and
H.~A.~Weidenm\"uller, Phys. Rep. {\bf 299},
189 (1998).

\bibitem{Yacoby} A.~Yacoby {\it et al.}, 
Phys.~Rev.~Lett., {\bf 74}, 4047 (1995).

\bibitem{Heiblum} 
E.~Schuster {\it et al.},
Nature, {\bf 385}, 417 (1997);

\bibitem{Buks} E.~Buks {\it et al.}, Physica, {\bf 249-251}, 295,
1998; D.~Sprinzak and M.~Heiblum, unpublished.

\bibitem{Deo} P.S.Deo and A.M.Jayannavar, Mod. Phys. Lett. {\bf
10}, 787 (1996)

\bibitem{Fano} H.~Xu and W.~Sheng,
Phys.~Rev. {\bf B 57}, 11903 
(1998);  C.-M.~Ryu and S.~Y.~Cho,
Phys.~Rev. {\bf B 58}, 3572 
(1998); H.-W.~Lee, Phys. Rev. Lett. {\bf 82}, 2358 (1999).

\bibitem{Oreg}  Y.~Oreg and Y.~Gefen, Phys. Rev. {\bf B 55}, 
13726 (1997).

\bibitem{Hackenbroich} G.~Hackenbroich, W.~D.~Heiss and
H.~A.~Weidenm\"uller, Phys. Rev. Lett. {\bf 79}, 127 (1997).

\bibitem{Gefen} R.~Baltin {\it et al.},
cond-mat/9807286.

\bibitem{Baltin} R.~Baltin and Y.~Gefen,
cond-mat/9907205.

\bibitem{CB} {\it Single charge tunneling: Coulomb
blockade phenomena in nanostructures, NATO ASI series}, edited
by H.~Grabert and M.~H.~Devoret (Plenum press 1992).

\bibitem{Matveev} K.~A.~Matveev, Sov. Phys. JETP {\bf 72}, 892
(1991); Phys. Rev. {\bf B 51}, 1743 (1995).

\bibitem{Leo} S.~Tarucha {\it et al.}, Phys. Rev. Lett. {\bf 77},
3613 (1996).

\bibitem{Bohr} A.~Bohr and B.~R.~Mottelson, {\it Nuclear Structure}, 
Benjamin, New York {\bf 1}, 284 (1969).

\bibitem{Landau}  L.D. Landau and E.M. Lifschitz {\it Quantum Mechanics},
p. 555, Pergamon, Oxford (1976). 

\bibitem{Glazman} L.~I.~Glazman and M.~E.~Raikh, JETP Lett. {\bf 47},
452 (1988)

\bibitem{Ng} 
T.~K.~Ng and P.~A.~Lee, Phys. Rev. Lett. {\bf 61}, 1768
(1988)

\bibitem{Wingreen} 
N.~S.~Wingreen and Y.~Meir, Phys.~Rev. {\bf B
49}, 11040 (1994).

\bibitem{Stopa} M.~Stopa, Physica {\bf B 251}, 228 (1998). 

\bibitem{preprint} P.~G.~Silvestrov and Y.~Imry, cond-mat/9903299


\bibitem{Kondo1} D.~Goldhaber-Gordon {\it et al.}, Nature {\bf
391}, 156 (1998); 
%D.~Goldhaber-Gordon {\it et al.}, 
Phys. Rev. Lett. {\bf 81}, 5225 (1998).

\bibitem{Kondo2} S.~M.~Cronenwett, T.~H.~Osterkamp, and 
L.~P.~Kouwenhoven, Science {\bf 281}, 540 (1998).

\bibitem{Kondo3} J.~Schmid {\it et al.}, Physica B {\bf 256-258}, 182
(1998).

\bibitem{Kondo4} F.~Simmel {\it et al.}, Phys. Rev. Lett. {\bf 83}, 
804 (1999).

\bibitem{Haldane} F.~D.~M.~Haldane, Phys. Rev. Lett. {\bf 40}, 
416 (1979).

\end{thebibliography}
\end{document}